\documentclass[doublecol]{epl2} 
\usepackage{amsmath,amssymb}
\usepackage{graphics,graphicx,color}
\usepackage{dcolumn,bm}
\usepackage{hyperref}
\usepackage{empheq}
\usepackage{amsfonts}
\usepackage{amsthm}
\usepackage{amssymb}
\usepackage{soul}
\usepackage{url}
\usepackage{caption}
\usepackage{subcaption}

\graphicspath{ {./figures/} }

\newcommand\norm[1]{\left\lVert#1\right\rVert}

\newcommand{\ncd}{\newcommand}
\ncd{\mrm}    {\mathrm}
\ncd{\beq}{\begin{equation}}
\ncd{\eeq}{\end{equation}}

\def\d{{\rm d}}

\ncd{\lamu}{\lambda_{\rm u}}
\ncd{\lamg}{\lambda_{\rm g}}

\begin{document}
\title{Localization and top eigenvalue detection}

\author{Diego Tapias\inst{1}, Benedikt Gr\"uger\inst{1}, Reimer K\"uhn\inst{2} \and Peter Sollich\inst{1,2} } 


\institute{                    
  \inst{1} Institute for Theoretical Physics, Georg-August-Universit\"at G\"ottingen, 37077 G\"ottingen, Germany \\
  \inst{2} Department of Mathematics, King's College London, London WC2R 2LS, UK
}

\abstract{The detection of the top eigenvalue and its corresponding eigenvector in ensembles of random matrices has significant applications across various fields. An existing method, based on the linear stability of a complementary set of cavity equations, has been successful in identifying the top eigenvalue when the associated eigenvector is extended. However, this approach fails when the eigenvector is localized. In this work, we adapt the real-valued cavity method to address this limitation by introducing a novel criterion that exploits the constraints of the cavity equations to detect the top eigenvalue in systems with a localized top eigenvector. Our results are validated using the Anderson model as a paradigmatic example.}

\maketitle

\section{Introduction}

The detection of the top eigenvalue and the corresponding top eigenvector of an ensemble of random matrices is a problem whose solution embraces many different applications~\cite{derrida1998exact, coja2006spectral, langville2006google, restrepo2008weighted, monasson2015estimating, moran2019may, jack2020ergodicity}. Techniques from the realm of disordered systems, such as replica theory and the cavity method,  have been successfully adapted to this problem for a number of different random matrix ensembles~\cite{kabashima2010cavity, kabashima2012first, susca2019top, budnick2025top}. In particular, an existing method~\cite{susca2019top} based on the cavity equations works well whenever the top eigenvector is extended. This is the case, for instance, of weighted adjacency matrices of sparse graphs, Markov transition matrices in a finite-dimensional state space, or invariant ensembles (see Refs.~\cite{susca2019top, susca2021statistical} for an extensive discussion and illustration of some of these systems). 

In this work, we adapt the existing method to find the top eigenvalue of systems with a localized top eigenvector. This is the situation in  many different disordered systems, with the Anderson model for electron transport in a disordered medium the paradigmatic example~\cite{anderson1958absence, abou1973selfconsistent}. In particular, any random matrix ensemble with  \emph{mobility edges}, i.e.\ values that separate extended from localized eigenvectors in the bulk of the spectrum~\cite{mott1967electrons, mott1987mobility, metz2010localization}), will exhibit a localized top eigenvector. An obvious exception occurs when the top eigenvalue is an outlier, i.e.\ separated from the bulk of the spectrum: there is then no general principle that tells us what the nature of the top eigenvector is. 

The current formulation of the cavity equations together with the criterion for the top eigenpair detection, which we refer to as the \emph{real--valued cavity method}, was developed in reference~\cite{susca2019top} inspired by earlier work by Kabashima~\cite{kabashima2010cavity, kabashima2012first}. It consists of the standard cavity equations (as presented for instance in Ref.~\cite{susca2021cavity}) without an imaginary regularizer, and focusses in these from the start on real (rather than complex) variables. This is equivalent, in the limit of infinite system size, to what has been dubbed the ``inverted thermodynamic limit''~\cite{kravtsov2018non, parisi2019anderson}. In addition to these equations, the real--valued cavity method comprises a linear system for an auxiliary set of fields, whose stability determines the actual condition for the detection of the top eigenvalue. In this work, we show that this stability criterion is meaningless in the case of a top localized eigenvector, because it remains satisfied in a regime in which the cavity variances are already negative and in consequence unphysical. 

The paper is organized as follows. First, we review the essence of the real--valued cavity method as developed in Ref.~\cite{susca2019top} and introduce our new criterion for the detection of the top eigenvalue for systems with a top localized eigenvector. Then, we illustrate the failure of the existing criterion and the plausibility of the new one, using as an example the Anderson model on a random graph. Then we show how both criteria compare in the case of finite instances of the Anderson model. Afterwards, we analyse the finite (population) size effects present in our method. Finally, we present our conclusions and provide a brief outlook towards future research.

\section{Real--valued cavity method}

Here we provide the minimal ingredients to obtain the system of equations relevant for our analysis. For a detailed derivation and discussion we refer the reader to Ref.~\cite{susca2019top}.

Let us consider an ensemble of sparse real--valued symmetric random matrices $\{\mathbf{J}\}$ (in general with non--zero diagonal) of dimension $N \times N$ each. Associated with a given realization, we construct the following ``grandcanonical'' measure on an adjoint space
\begin{align}
  P_{\beta, J} (v_1, \ldots, v_N) = \frac{1}{Z} \exp \left( \frac{\beta}{2} \bigg( \sum_{i,j= 1}^N v_i J_{ij} v_j - \lambda \sum_{i = 1}^N v_i^2 \bigg) \right)
  \label{jointvs}
\end{align}
Here $\lambda$ is an auxiliary Lagrange multiplier and the partition function is given by
\begin{align}
  Z = \int \d v_1 \cdots \d v_N  \exp \left( \frac{\beta}{2} \bigg( \sum_{i,j= 1}^N v_i J_{ij} v_j - \lambda \sum_{i = 1}^N v_i^2 \bigg) \right) \, .
  \label{part} 
\end{align}
It turns out that the joint distribution of the top eigenvector entries is the limit $\beta \to \infty$ of equation~\ref{jointvs} evaluated exactly at the top eigenvalue, $\lambda = \lambda_1$. Thus, in order to find the top eigenvector we need first to obtain the corresponding eigenvalue. In what follows we derive the cavity equations associated with the measure~\ref{jointvs} and introduce the criterion to find the top eigenvalue $\lambda_1$.

To each random matrix $\{\mathbf{J}\}$ we can associate a graph with nodes labelled by $i$ and a set $\mathcal{E}$ of edges $(i,j)$ identified by the nonzero off-diagonal matrix elements $J_{ij}$. If the matrices are suitably sparse we can assume that the corresponding graphs are locally tree--like.  
The marginal distribution on site $i$ can then be written in the general form
\begin{align}
  P_i(v_i) \propto &\exp\left(  \frac{\beta}{2} (J_{ii} - \lambda) v_i^2 \right) \times \notag \\
  &\prod_{j \in \partial i} \int \d v_j \, {\mathrm{e}}^{\beta J_{ij} v_i v_j} P_j^{(i)} (v_j)
    \label{marginal}
\end{align}
where $P_j^{(i)} (v_j)$ is the cavity distribution on site $j$, i.e.\ the distribution obtained after removing node $i$ and all corresponding edges (matrix elements) $J_{ij}$.
The notation $\partial i$ refers to all the neighbours of node $i$. This cavity distribution can be written in terms of analogous cavity distributions resulting from removal of the neighbour nodes $j$ themselves:
\begin{align}
  P_j^{(i)}(v_j) \propto  &\exp\left( \frac{\beta}{2} (J_{jj} - \lambda) v_j^2 \right) \times \notag \\
  &\prod_{l \in \partial j \backslash i} \int \d v_l\,  {\mathrm{e}}^{\beta J_{jl} v_j v_l} P_l^{(j)} (v_l)
  \label{caveq}
\end{align}
where the set notation $\partial j \backslash i$ means all the neighbours of node $j$ except $i$. Equation~\eqref{caveq} is a self--consistent set of equations for the cavity distributions that can be solved by a Gaussian \emph{ansatz}
\begin{align}
 P_j^{(i)}(v_j) \propto \exp\left(-\beta \frac{{H_j^{(i)}}^2}{2 \Omega_j^{(i)}}\right)
\exp\left(-\frac{\beta}{2}\Omega_j^{(i)} v_j^2 + \beta H_j^{(i)} v_j\right) \,.
  \label{cav_ansatz}
\end{align}
Here $ \Omega_j^{(i)}$ is the inverse variance of the Gaussian, known in statistics as the precision, so we call 
the $ \Omega_j^{(i)}$ cavity precisions, while for the $ H_{j}^{(i)}$ we use the name auxiliary cavity fields. Inserting equation~\eqref{cav_ansatz} into equation~\eqref{caveq}, one obtains the following self--consistent equations for the cavity fields
\begin{align}
  \Omega_j^{(i)} &= \lambda - J_{jj} - \sum_{l \in \partial j \backslash i} \frac{J_{jl}^2}{\Omega_l^{(j)}} \label{cav1} \\
  H_j^{(i)} &= \sum_{l \in \partial j \backslash i} \frac{J_{jl}}{\Omega_l^{(j)}} H_l^{(j)} \label{cav2}
\end{align}

The condition for the top eigenvalue given in Ref.~\cite{susca2019top}, is based on the stability of the linear equations~\eqref{cav2}.  Given the fixed--point solution of eq.~\eqref{cav1} for the cavity precisions let us construct the non--backtracking operator
\begin{equation}
  B_{(i,j),(k,l)} =  \begin{cases}
    \frac{J_{jl}}{\Omega_l^{(j)}} \quad  i \neq l \wedge j = k \\
    0 \quad \mathrm{otherwise}
  \end{cases}
\end{equation}
 Additionally, let us collect the auxiliary cavity fields into a single $2M$--dimensional vector, with $M$ the number of edges of the network; then after relabelling any directed edge $(i,j)$ with a new single index $a$ so that $H_a\equiv H^{(i)}_j$, equation~\eqref{cav2} reads as
\begin{align}
  H_a = \sum_{b = 1}^{2M} B_{ab} H_b
  \label{linearH}
\end{align}
If we start from some set of initial guesses $\bm{H}_0$, a solution of eq.~\eqref{cav2} can thus be approached by  repeatedly acting on $\bm{H}$ with the operator $\mathbf{B}$. We write this iteration as
\begin{align}
  \bm{H}_t = \mathbf{B} \bm{H}_{t-1}
  \label{updatehs}
\end{align}
with $t$ the label of each iteration and define the growth rate $\eta$ as the large $t$--limit of the ratio of the norms of the vector $\bm{H}$ after one iteration, i.e.
\begin{align}
  \eta = \lim_{t \to \infty} \frac{\norm{{\bm{H}}_t}}{\norm{\bm{H}_{t-1}}}
  \label{growth}
\end{align} 
It can then be shown (see Ref.~\cite{kabashima2012first} for a proof) that eq.~\eqref{cav2} has a non--trivial solution if and only if $\eta = 1$, which happens exactly at $\lambda = \lambda_1$. This is the basis for what we will call the \emph{growth--rate} criterion, which estimates $\lambda_1$ as the value $\lamg$ that yields a unit growth rate.

An alternative criterion, which we propose here for localized eigenvectors, is based on a constraint on the cavity precisions, namely, the fact that all of them have to be positive to have a well defined measure. This criterion is based on the finding that solving  equation~\eqref{cav1} for $\lambda < \lambda_1$ yields at least one negative cavity precision whenever the top eigenvector is localized. On this basis we introduce the following estimator
\begin{align}
  \lamu = \min \{\lambda \in \mathbb{R} :  \Omega_j^{(i)} (\lambda)  > 0 \, \forall \, (i,j) \in \mathcal{E} \}
  \label{new_si}
\end{align}
where $\mathcal{E}$ is the set of edges of the network. In words, $\lamu$ is the minimum of the $\lambda$--values that yield physical (positive) cavity precisions. Numerically, one can then use e.g.\ bisection to estimate the top eigenvalue using the criterion stated in equation~\eqref{new_si}.

In the thermodynamic limit, $N \to \infty$, the cavity equations translate into a self–consistency equation for the joint distribution of  cavity  precisions and auxiliary fields, $q(\omega, h)$, which can be solved using a standard Monte Carlo method known as population dynamics~\cite{susca2021cavity}. In that case, we reinterpret equation~\eqref{updatehs} as the change in the population of the $h$-fields ($\{h_i\}_{1\leq i \leq N_P}$, with $N_P$ the population size) after one sweep, the latter being an iteration in which each member of the population has been updated once \emph{on average}. The growth rate (eq.~\eqref{growth}) is then estimated from the change in the norm of the auxiliary fields after many sweeps. On the other hand, the alternative criterion~\eqref{new_si} reads, in the population dynamics framework, as
\begin{align}
  \lamu = \min \{\lambda \in \mathbb{R} : \omega_i(\lambda)  > 0 \, \forall \, i  \in N_{P} \}
  \label{new_np}
\end{align}
where $\{\omega_i \}_{1\leq i \leq N_P}$ is the population of cavity precisions.

    \begin{figure}
        \centering
        \includegraphics[width=0.45\textwidth]{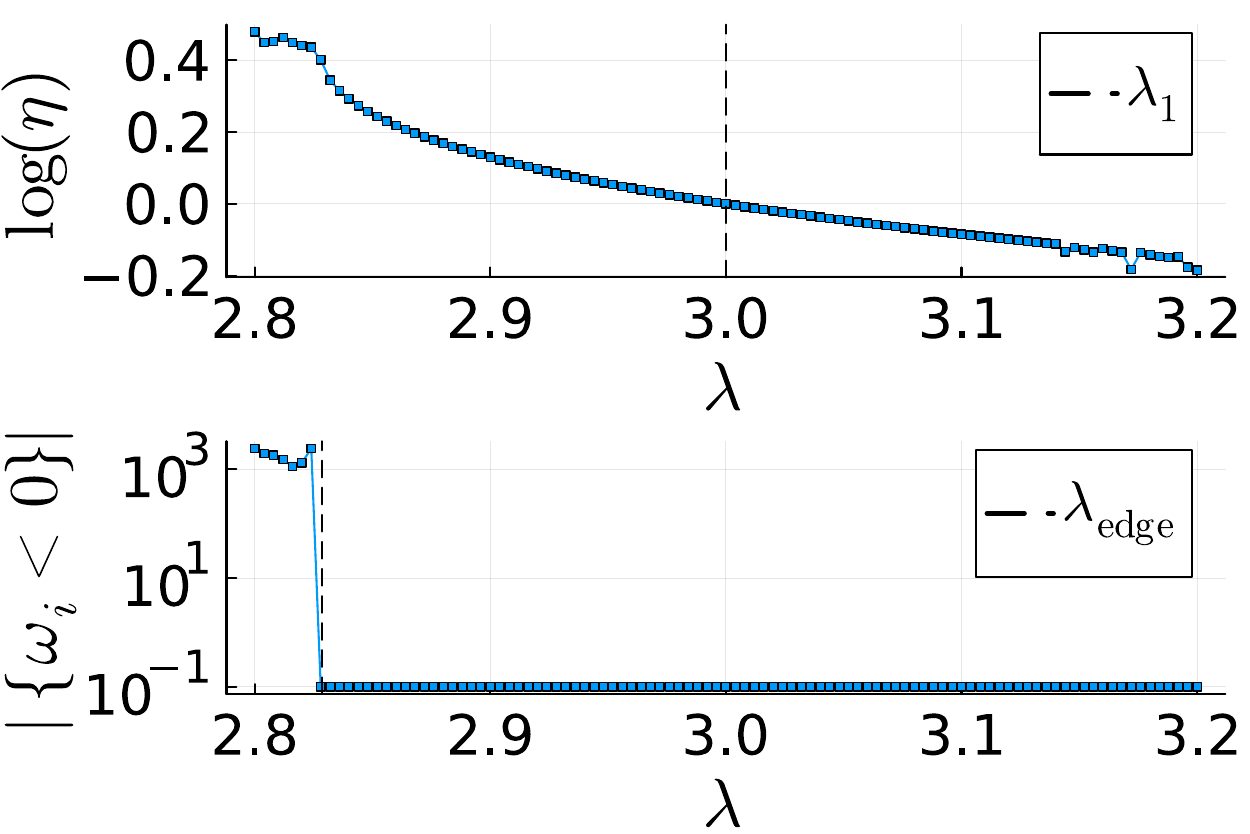}
            \caption{Logarithm of the growth rate (eq.~\eqref{growth}, top panel) and number of negative cavity precisions (plus a small $\delta = 10^{-1}$ to help visualization in logarithmic scale, bottom panel) against $\lambda$, from population dynamics for the set of adjacency matrices of a random regular graph with connectivity $c = 3$. The estimated top eigenvalue based on the growth rate criterion, $\eta = 1$, is the dashed line in the top panel. In the bottom panel, the spectral edge corresponds to the dashed line.  Population size: $N_P = 10^5$.}
         \label{fig:pd_extended}
       \end{figure}
      \begin{figure}
        \centering
        \includegraphics[width=0.45\textwidth]{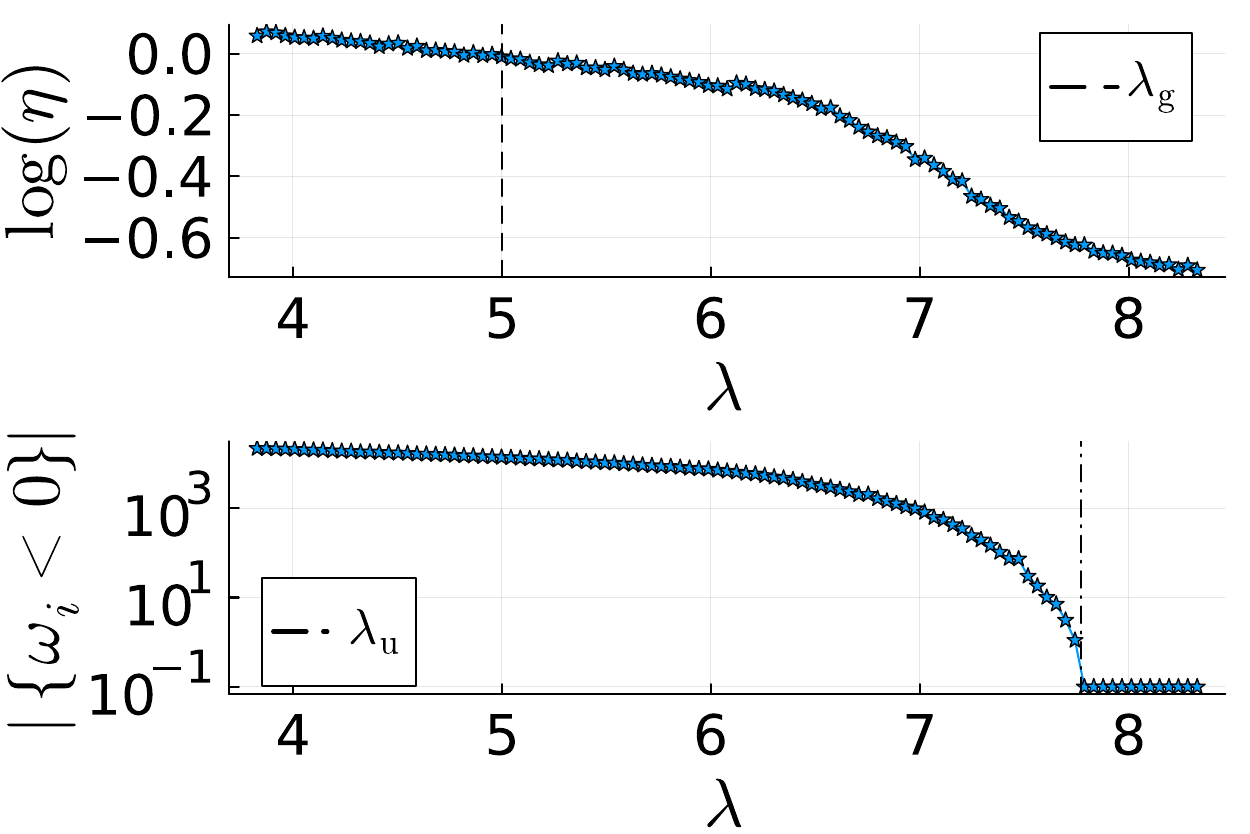}
            \caption{Logarithm of the growth rate (top panel) and number of negative cavity precisions (bottom panel) against $\lambda$, from population dynamics for the Anderson model on a random regular graph. The estimated top eigenvalue based on the growth rate criterion, $\eta = 1$ is the dashed line in the top panel. In the bottom panel, the estimator based on relation~\eqref{new_np} corresponds to the dashed line. Population size: $N_P = 10^5$. Parameters of the model: $W = 12$, $c = 3$. The theoretical top eigenvalue for the parameters used is $\lambda_1 \approx 8.83$. The proposed new criterion~\eqref{new_np} provides an estimate much closer to this than the one based on the growth rate, though it does remain affected 
            by strong finite (population) size effects as
shown in Fig.~\ref{fig:scaling}.
            }
         \label{fig:pd_localized}
       \end{figure}

\section{Population dynamics for the detection of the top eigenvalue}

In Fig.~\ref{fig:pd_extended} we show the behavior of the growth rate across a $\lambda$--grid around the top eigenvalue for the ensemble of adjacency matrices of a random regular graph, with connectivity $c$. In this case, the top eigenvector is extended (see Ref.~\cite{huang2024spectrum} for rigorous proofs of all graph theoretical statements in this paragraph) and the estimator $\lamg$ is accurate, as the known top eigenvalue is $\lambda_1 = c$. We point out that for this model, this top eigenvalue is an outlier, i.e.\ there is a spectral gap separating it from the bulk of eigenvalues. The spectral edge, meaning the maximum value of $\lambda$ for which the spectral density is non--vanishing, is located at $\lambda_{\rm{edge}} = 2 \sqrt{c-1}$. In Fig.~\ref{fig:pd_extended} we also show that around this value, negative (unphysical) cavity precisions appear. This observation is interesting as for the models that we analyse below, the localized top eigenvector coincides with the spectral edge.

In Fig.~\ref{fig:pd_localized} we show plots analogous to those in Fig.~\ref{fig:pd_extended}, but now for the Anderson model on a random regular graph, with again connectivity $c=3$. In the appendix we recall the main properties of the Anderson model on a random regular graph. Its main feature for the purposes of the present discussion is that  there is no spectral gap and the top eigenvector is always localized. From the figure, we notice that $\lamg$ is clearly unsuitable as an estimator of the top eigenvalue, given that it lies in a $\lambda$--range where the cavity precisions are already unphysical. Additionally, the growth--rate estimator considerably underestimates the theoretical top eigenvalue~\cite{abou1973selfconsistent, biroli2010anderson}, i.e.\ $\lamg \ll \lamu < \lambda_1 =   2 \sqrt{c - 1} + \frac{W}{2}$, where $W$ is the parameter that controls the distribution of the on--site energies in the Anderson model. The cavity precision criterion~\ref{new_np} yields an estimate distinctly closer to the theoretical value but still somewhat lower; this is due to finite size effects as discussed below.

\section{Top eigenvalue detection for single instances with a localized eigenvector}

\begin{figure}
        \centering
        \includegraphics[width=0.49\textwidth]{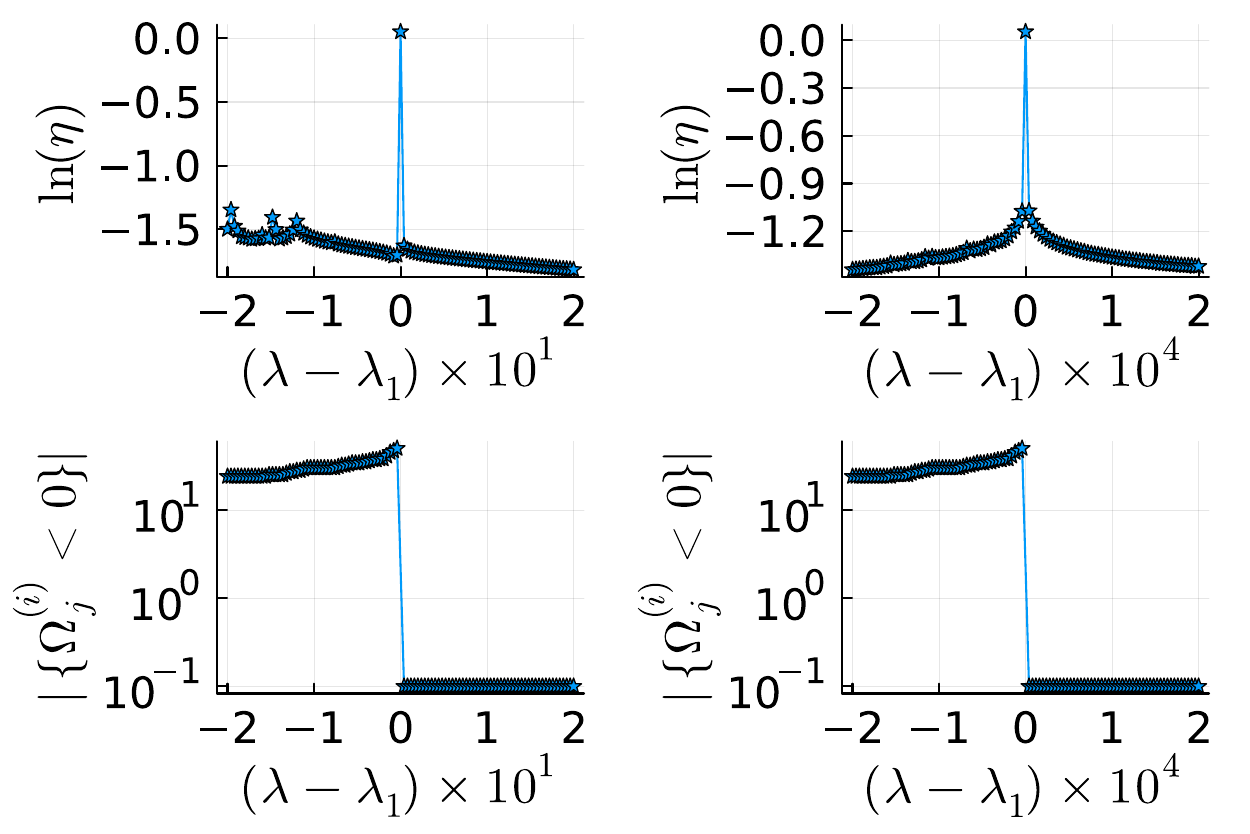}
            \caption{Logarithm of the growth rate (top panel) and  number of negative cavity precisions  (bottom panel) against $\lambda - \lambda_1$ for an instance of the Anderson model on a random regular graph with connectivity $c=3$, size $N = 2^{15}$ and disorder strength $W = 12$. The top eigenvalue $\lambda_1$ is obtained by exact diagonalization. The two columns show two different resolutions for the $x-$axis.}
         \label{fig:si_panel}
       \end{figure}

While we have focussed on population dynamics results above, there are issues with the growth--rate estimator already at the level of single random matrix instances (finite $N$) when the top eigenvector is localized. In Fig.~\ref{fig:si_panel} we show the typical behavior of the relevant probes around the top eigenvalue, again for the Anderson model on a finite random regular graph. From the figure, one sees that the growth rate exhibits a very narrow peak close to $\lambda_1$. Hence, while the growth--rate criterion remains applicable in principle and does give the correct $\lambda_1$, in practice it becomes unreliable as even on a relatively fine numerical grid the narrow peak in the growth rate is easy to miss. One interpretation of the population dynamics results shown above is then that the effective average over the random matrix ensemble implied by the population dynamics setting ``washes out'' the peak seen for single random matrix instances, leading to the spurious estimate of a much lower $\lambda_1$. 
In contrast the estimator~\ref{new_si} is robust also for single instances (Fig.~\ref{fig:si_panel}): the top eigenvalue is enclosed (with high accuracy) in the $\lambda$ interval that contains the change from physical to unphysical cavity precisions.

       \section{Finite size (and population size) effects}

For the Anderson model, the existence of Lifshitz tails makes the estimation of the spectral edge difficult~\cite{lifshitz1964energy, biroli2010anderson}. Therefore we expect the presence of strong finite size effects for single instances, or population size effects for the thermodynamic limit solved via population dynamics.

\begin{figure}
        \centering
        \includegraphics[width=0.45\textwidth]{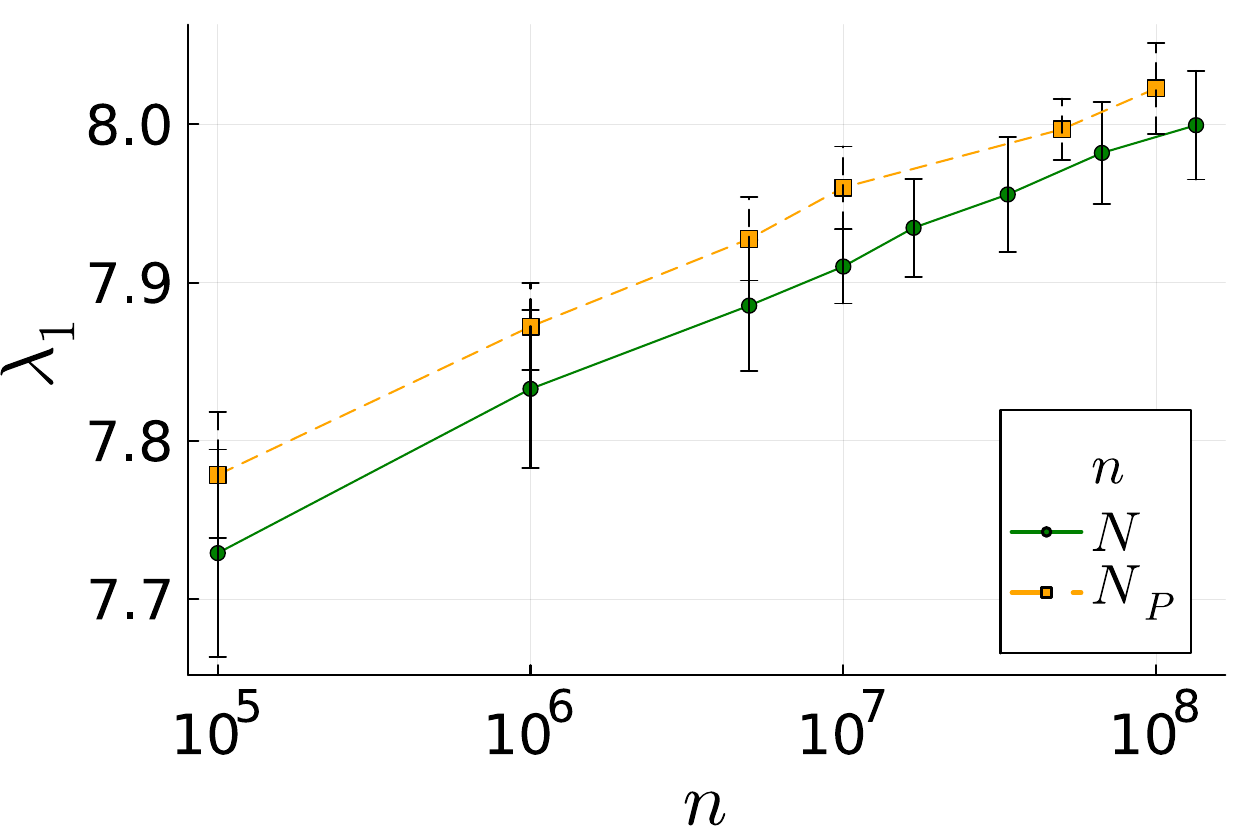}
            \caption{Dependence of the top eigenvalue with system size ($N$) and population size ($N_p$). Error bars show the standard deviation of statistics over $2^{30}/N$  instances and $10^9/N_P$ runs, respectively. Parameters of the model: $W = 12$, $c = 3$.}
         \label{fig:scaling}
       \end{figure}

       We take advantage of the sparsity of the system to obtain the top eigenvector via exact diagonalization (using the Lanczos algorithm as implemented in the \emph{Arpack.jl} library~\cite{arpack}) of single instances with sizes up to $N = 2^{27}$. In figure~\ref{fig:scaling} we show the scaling of $\lambda_1$ with $N$ for a wide range of system sizes. We fit this to the function $\lambda_1 = c_0 + c_1 N^{-\alpha_1}$, with $c_0 = \hat{\lambda}_1^{N\to \infty}$, i.e. the estimator of the top eigenvalue for the thermodynamic limit. The result $ \hat{\lambda}_1 ^{N\to \infty} \approx 8.6 \pm 0.3$ still has a substantial error bar, but does cover within the one-standard deviation range the theoretical result of $\approx 8.83$ for the parameters used. We display in addition the scaling of the estimator $\lamu$ obtained from population dynamics plus  bisection, as a function of the population size $N_P$. It is remarkable that the size scaling from population dynamics is essentially the same as that from exact diagonalization. This tells us, on the one hand, that finite {\em population} size effects are of the same order as finite {\em instance} size effects for this model, implying that the advantages of using population dynamics for the study of the thermodynamic limit are limited. On the other hand, the observed scaling validates our estimator based on the appearance of unphysical cavity precisions (equation~\eqref{new_np}): we observe behavior that is qualitatively equivalent to the one found by exact diagonalization, where  no assumption or approximation is made in the determination of the top eigenvalue.
       
       \section{Conclusions}

       We have shown that the detection of the top eigenvalue with a corresponding localized eigenvector using a population dynamics algorithm is unreliable using the existing growth--rate criterion. For single instances, this is due to the fact that the growth--rate exhibits only a very narrow peak (approaching unity) in the immediate vicinity of the top eigenvalue, making its detection very difficult. We therefore proposed a new criterion that looks for the appearance of unphysical (negative) cavity precisions as $\lambda$ is decreased below $\lambda_1$. We showed for the case of the Anderson model on a random regular graph that the criterion is reliable for the estimation of the top eigenvalue in the thermodynamic limit.

       A puzzling problem that we have left out from our analysis here, and which requires further investigation, is the analysis of the distribution of the components of the top {\em eigenvector} when the latter is localized. In the case of single instances it turns out that predictions for this distribution are, like the growth--rate, quite sensitive to the value of $\lambda$. For population dynamics the results are more stable according to our preliminary investigations, but the distribution of the entries does not follow the prediction from theory (in essence equation~\eqref{jointvs} for $\beta \to \infty$). 
       Further research towards a robust method for predicting the distribution of top eigenvector components is therefore an interesting avenue for the future.

      A direct application of our work is the detection of the dynamical free energy for biased stochastic systems~\cite{jack2020ergodicity}. For those systems that can be formulated in terms of a biased master operator, the top eigenvalue is nothing but their free energy (see Ref.~\cite{jack2020ergodicity} and references therein). Reliably obtaining this quantity is crucial for the detection of dynamical phases. Work is in progress in this direction for the biased (sparse) Bouchaud trap model, which in contrast to its mean field counterpart~\cite{tapias2024bringing} is not solvable in closed form.


       \section{Appendix}

\subsection{Anderson model on a RRG}

A random regular graph, or Bethe lattice, is a random network in which every node has the same number of neighbours. This number is the (mean) connectivity $c$ of the network. From the random matrix theory perspective, the Anderson model on a random graph is defined as the set of matrices $\{ \bf{J} \}$ of size $N \times N$ such that a given instance is characterized by the following elements
\begin{align}
  J_{ij} = E_i \delta_{ij} - A_{ij}
  \label{anderson_m}
\end{align}
where the matrix $\bf{A}$ is the adjacency matrix of the network and the diagonal elements (on--site energies) $\{E_i \}$ are i.i.d.~random variables uniformly distributed in the interval $[-W/2, W/2]$. Thus for a fixed connectivity, the disorder strength $W$ is the key control parameter.

An analytical treatment of this model~\cite{acosta1992analyticity, klein1998extended} yields the following prediction for the top edge of the spectrum 
\begin{align}
  \lambda_1 = 2 \sqrt{c - 1} + \frac{W}{2}
  \label{theo_top}
\end{align}
However, the numerical estimation of $\lambda_1$ for $N\to\infty$ is a difficult problem, as the density of states decays faster than an exponential at the edges and therefore becomes extremely small (Lifshitz tail)~\cite{lifshitz1964energy, biroli2010anderson}.

Finally, the equations (\eqref{cav1} and~\eqref{cav2}) for the cavity precisions and auxiliary fields for an Anderson matrix (eq.~\eqref{anderson_m}) read as follows:
\begin{align}
    \Omega_j^{(i)} &= \lambda - E_j - \sum_{l \in \partial j \backslash i} \frac{1}{\Omega_l^{(j)}} \label{cav1_and} \\
  H_j^{(i)} &= \sum_{l \in \partial j \backslash i} \frac{H_l^{(j)}}{\Omega_l^{(j)}}  \label{cav2_and}
\end{align}
The system of equations~\eqref{cav1_and} was analysed in detail in Ref.~\cite{parisi2019anderson} in terms of the propagators: $G_j^{(i)} = \frac{1}{\Omega_j^{(j)}}$, in order to derive a criterion for the localization--delocalization transition based on susceptibilities.

\bibliographystyle{unsrt}
\bibliography{bibliography}

\end{document}